# Electron-phonon coupling reflecting dynamic charge inhomogeneity in copper-oxide superconductors

D. Reznik[1,2], L. Pintschovius[1], M. Ito[3], S. Iikubo[3], M. Sato[3], H. Goka[4*], M. Fujita[4], K. Yamada[4], G.D. Gu[5], and J.M. Tranquada[5]

*1 Forschungszentrum Karlsruhe, Institut für Festkörperphysik, P.O.B. 3640, D-76021 Karlsruhe, Germany*

*2 Laboratoire Léon Brillouin, CE Saclay, F-91191 Gif-sur-Yvette Cedex, France*

*3 Department of Physics, Division of Materials Science, Nagoya University, Furo-cho, Chikusa-ku, Nagoya 464-8602, Japan*

*4 Institute for Material Research, Tohoku University, Katahira, Aoba-ku, Sendai, 980-8577, Japan.*

*5 Condensed Matter Physics and Materials Science Department, Brookhaven National Laboratory, Upton, New York 11973-5000, USA*

**The attempt to understand cuprate superconductors is complicated by the presence of multiple strong interactions. While many believe that antiferromagnetism is important for the superconductivity, there has been revived interest in the role of electron-lattice coupling.**[1,2,3,4] **The recently studied conventional superconductor $MgB_2$ has a very strong electron-lattice coupling, involving a particular vibrational mode (phonon), that was predicted by standard theory and confirmed quantitatively by experiment.**[5] **Here we present inelastic scattering measurements that show a similarly strong anomaly in the Cu-O bond-stretching phonon in the cuprate superconductors $La_{2-x}Sr_xCuO_4$ (with x=0.07, 0.15). This is in contrast to conventional theory, which does not predict such**

**behavior. The anomaly is strongest in $La_{1.875}Ba_{0.125}CuO_4$ and $La_{1.48}Nd_{0.4}Sr_{0.12}CuO_4$, compounds that exhibit spatially modulated charge and magnetic order, often called stripe order.[6] It occurs at a wave vector corresponding to the charge order. These results suggest that this giant electron-phonon anomaly, which is absent in undoped and over-doped non-superconductors, is associated with charge inhomogeneity. It follows that electron-phonon coupling may be important to our understanding of superconductivity, although its contribution to the mechanism is likely to be indirect.**

Ordinary metals can be understood within the conventional Fermi-liquid picture where electron-like quasiparticles populate bands in energy-momentum space up to the cutoff at the Fermi energy $E_f$. The dispersion of these bands is determined by the crystal structure and chemistry. Lattice vibrations couple to electrons because displacements of atoms from their equilibrium positions alter the band dispersions, lowering or raising the total electronic energy. Calculations based on this standard model have been very successful in explaining many properties associated with electron-phonon coupling in conventional metals, especially including phonon-mediated superconductivity. In the case of copper oxide superconductors, however, such electron-phonon interactions appear to be too weak to explain the high superconducting transition temperatures ($T_c$). Furthermore, the applicability of the Fermi liquid model to the copper oxides has been questioned.

In one alternative approach, it has been argued that the competition between the kinetic energy of charge carriers and the antiferromagnetic superexchange between magnetic moments on neighboring Cu atoms leads to charge segregation.[7,8] Experimental support for such a view was provided when static charge segregation in the form of stripes was first discovered in $La_{1.48}Nd_{0.4}Sr_{0.12}CuO_4$,[6] where a small structural modification provides a pinning potential for the stripes. The fact that static stripes compete with superconductivity has led many to question the relevance of charge inhomogeneity to understanding the superconductivity. The possible existence of dynamic charge inhomogeneity has been difficult to establish in the absence of a clear experimental signature.[9]

Charge inhomogeneity with a periodic modulation should affect interatomic Coulomb forces and soften lattice vibrations at the same wave vector. For example, in compounds with stripes one expects naively that the strongest effect should be in the longitudinal bond-stretching mode whose polarization pattern matches the lattice deformation induced by the charge inhomogeneity (see Fig. 1). This idea has motivated us to test whether some feature of the relevant phonon mode might provide a useful measure of dynamic charge inhomogeneity. While there have been previous reports of phonon anomalies in copper-oxide superconductors[10,11,12], and unexpected phonon dispersions have been observed in $La_{1.69}Sr_{0.31}NiO_4$ with diagonal stripe order[13], no one has made the crucial test of comparing electron-phonon coupling in copper oxide compounds at doping levels associated variously with static stripe order, superconductivity, and Fermi liquid behavior.

Thus, we began our study by investigating the bond-stretching vibration along the Cu-O bond direction in $La_{1.48}Nd_{0.4}Sr_{0.12}CuO_4$[14] and $La_{1.875}Ba_{0.125}CuO_4$,[15] compounds in which diffraction measurements have demonstrated the presence of static stripes. In addition, a soft-x-ray resonant-diffraction study[16] of $La_{1.875}Ba_{0.125}CuO_4$ has recently provided direct evidence for concomitant charge order, below $T_s \approx 60K$, with the modulation wave vector $\mathbf{q}_{co}$=(0.25 0 0). We performed most of our measurements at low temperature, $T \approx 10$ K, in the ordered stripe phase.

Figure 2(a) shows a color-coded map of the bond-stretching phonon spectra in $La_{1.875}Ba_{0.125}CuO_4$ at T = 10 K for phonon wave vectors $\mathbf{q}$ = (h 0 0). It is immediately clear that the well-defined branch dispersing downward from the zone center acquires a broad low-energy tail near to $\mathbf{q}_{co}$. In that vicinity, the individual scans can be fit with a two-peak structure [see Fig. 3(a,b) and discussion below], with the line shape returning to a single peak at the zone boundary. The integrated intensity of the entire feature agrees very well with the calculated[17] structure factor of the bond-stretching mode, as shown in Fig. 2(b).

Figure 3 compares individual scans through the phonon mode at **q** = (0.15 0 0) and at $\mathbf{q}_{co}$; these points are separated by slightly more than the full width at half maximum (FWHM) of the instrumental **Q**-resolution. Whereas at **q** = (0.15 0 0) there is a single sharp peak, at $\mathbf{q}_{co}$ the main peak is much weaker and the missing intensity appears in the low energy tail. These latter spectra can be fit with two displaced peaks with different energy widths but comparable integrated intensities. Extending the two-peak analysis to the full **q** range (and assuming resolution-limited energy widths at small **q**) leads to a model of two phonon branches (instead of one), as indicated by the solid lines in Fig. 2(a). One branch has a monotonic, cosine-like downward dispersion in the (1 0 0) direction. The other branch is highly anomalous in that it drops abruptly by more than 10 meV at about $\mathbf{q}_{co}$ and then rises back up towards the zone boundary to merge with the other branch; it also has an anomalously large energy width. The inferred drop is much sharper than the instrumental resolution in the longitudinal direction, and also in the transverse direction (not shown), so that the average over wave vectors within the resolution ellipsoid produces the broad low-energy tail at $\mathbf{q_{co}}$. A splitting of the bond-stretching mode is compatible with the crystal symmetry in the ordered stripe phase. The inferred variation between the two branches could indicate a substantial difference for phonons propagating parallel and perpendicular to the stripes. (The symmetry of the crystal also ensures that we would average over both components.)

Consistent with the model of stripes aligned along Cu-O bonds, the dispersion along the [110] direction is much more normal. The inset of Fig. 2(a) shows that the bond-stretching branch in the [1 1 0] direction is completely flat. It has a broad linewidth, which may reflect bond stiffness variations due to the inhomogeneous charge distribution. Also, the anomalous lineshape along [100] is independent of the out-of-plane (0 0 1) component of **q**, in agreement with the stripe dynamics being uncorrelated between neighboring $CuO_2$ planes. Furthermore, the softening and broadening effects are strongest in the bond-stretching branch: detailed measurements did not find any anomalies in the bond-bending branch (appearing near 60 meV in Fig. 2a).

Temperature dependence provides further confirmation of the unusual nature of the phonon line shape near $q_{co}$. The anomaly is the strongest at the lowest temperature, 10K (Fig.3a,b). At 330K [see Fig. 3(c)] the low-energy tail is greatly reduced and the peak is enhanced, though even at this temperature the conventional line shape, as indicated by the curve at **q** = (0.15 0 0), is not fully recovered. The conventional anharmonic behavior of phonons is to *broaden* and *soften* with increasing temperature, not *sharpen* as we observe.

To test the doping dependence of the anomaly we have measured large single crystal samples of underdoped, $T_c$=20 K, optimally-doped, $T_c$=35K, and overdoped nonsuperconducting $La_{2-x}Sr_xCuO_4$ (x=0.07, 0.15, 0.3, respectively). Figure 4(a) compares low-temperature scans at **q** = (0.25 0 0) for all of these samples, together with a previous result for $La_2CuO_4$. As one can see, the peaks are considerably broader for x = 0.07 and 0.15 than for the undoped and over-doped samples. In Fig. 4(b) we compare the dispersion of the bond-stretching branch for these samples with published data on $YBa_2Cu_3O_{6.6}$ ($T_c$=66K)[18] and $YBa_2Cu_3O_{6.95}$ ($T_c$=93K)[19]. The behavior in the superconducting compounds deviates substantially from the simple cosine dispersion with narrow linewidths found for the over-doped nonsuperconducting $La_{1.7}Sr_{0.3}CuO_4$ with the biggest effects at **q**=(0.25-0.3 0 0) (Fig. 4(c)) We note that, consistent with the trend found here, a similar deviation from cosine-like dispersion, together with a large energy broadening, has been reported for superconducting $HgBa_2CuO_{4+\delta}$[20] at the same wave vectors. Thus, there seems to be a general correlation between the occurrence of strong electron-phonon coupling effects in the bond-stretching mode half way to the zone boundary and a hole concentration in the range compatible with superconductivity in the cuprates.

In terms of temperature dependence, the onset of the anomaly moves to lower temperatures with increasing x in both $La_{2-x}Sr_xCuO_4$ and $YBa_2Cu_3O_{6+x}$. For example, the anomaly is only slightly reduced in $YBa_2Cu_3O_{6.6}$ when the temperature is raised to 300K[18], but in $YBa_2Cu_3O_{6.95}$ the normal dispersion is already recovered by 200K[19,21]. The dispersions of the same phonon branch in the nonsuperconducting undoped $La_2CuO_4$ and overdoped $La_{1.7}Sr_{0.3}CuO_4$ do not show any sign of the anomaly at any temperature.

We believe that McQueeney et al.[10] and Chung et al.[11] reported manifestations of the same effect in $La_{1.85}Sr_{0.15}CuO_4$ and $YBa_2Cu_3O_{6.95}$, respectively, that are compatible with our observations; however, our measurements, performed with much higher resolution and better signal-to-background ratio, do not confirm their evidence for unit cell doubling. Compared with Ref. 22, the anomaly appears to be stronger in the present study possibly due to higher sample quality. Other previous reports of phonon anomalies in the copper oxides focused on an entirely different effect – softening of the so-called "half breathing" mode at the zone boundary, $\mathbf{q}=(0.5\ 0\ 0)$, with increased doping.[12] This effect is distinct from the one we report here since it continues into the overdoped regime[23]; in fact it is responsible for the downward cosine dispersion that we consider as "normal" behavior, since it is predicted by conventional calculations.[3]

In the case of superconducting $MgB_2$, one of the highest $T_c$ superconductors ($T_c$ = 39 K) for which electron-phonon coupling is agreed to be responsible for electron pairing, very large (10-15 meV full width at half maximum) phonon linewidths have been measured.[5] Both the linewidths and the dispersions are in good agreement with calculations based on the conventional theory of electronic structure (using the local-density approximation). The effect there is explained in terms of a Kohn anomaly.

The sharp dispersion and strong line broadening that we observe in the bond-stretching mode of superconducting cuprates also looks similar to a Kohn anomaly or charge-density-wave instability. So can these results be understood within conventional models that assume electronic homogeneity? The answer seems to be no. Several independent electron-phonon coupling calculations fail to predict the huge effect that is observed here.[2,3,4] They all give a very similar Fermi surface and adequately account for many nontrivial features of the phonon spectrum, such as the downward dispersion of the normal bond-stretching branch at optimal doping and the modest electron-phonon coupling for the bond-buckling branch dispersing from the $B_{1g}$ Raman mode near 42 meV; however, these calculations greatly underestimate the renormalization of the anomalous branch. For example, Refs. 2 and 3 give an electron-phonon coupling constant for the bond-stretching branch that is much smaller than that for the bond-buckling branch, which is renormalized by at most 1meV.[24]

Of course, strong electron-electron correlations are expected to be important in copper oxides for electronic states near the Fermi level, and these effects are not properly described in the conventional theory. An example of a model intended to capture correlation effects is the so-called t-J model. Extensions of that model to include electron-phonon coupling allow a reasonable description of the doping dependence of the half-breathing mode.[25] We note that Ishihara and Nagaosa[26] were able to get a substantial dip in the bond-stretching branch at $\mathbf{q} \approx (0.3\ 0\ 0)$; however, they had to use an off-diagonal electron-phonon coupling strength an order of magnitude larger than that estimated by Rösch and Gunnarsson.[25]

The empirical result found here is that the largest dispersion dip and line broadening occur in samples that exhibit static stripe order. Theoretically, Kaneshita et al.[27] have predicted that such behavior should occur at $\mathbf{q}_{co}$ for the bond-stretching mode propagating in a direction perpendicular to the stripes. Of course, the phonons that we have studied are at fairly high energy, and the response at such energies need not show great sensitivity to the presence or absence of correlations at zero energy (i.e., static order). In the superconducting copper oxides, we believe that the anomaly reflects a response to dynamic charge inhomogeneity. The orientation of the response with respect to the charge modulation wavevector is not clear cut. Also, based on the stripe picture the strongest anomaly in $La_{1.93}Sr_{.07}CuO_4$ should be at $k = 0.14$ (magnetic incommensurability is $0.07$[28]), and not at 0.3 as observed in the experiment (Fig. 3c). Furthermore, a stripe-based interpretation of the anisotropic magnetic response[29] measured in $YBa_2Cu_3O_{6+x}$ would suggest dynamic stripes oriented along the [010] direction, parallel to the Cu-O chains, and this is the same direction in which the large phonon anomaly appears. Thus, the possible role of Fermi-surface nesting effects associated with electronic states with wave vectors parallel to charge stripes should be considered.[30]

To conclude, our experiments show strong bond-stretching phonon anomalies common to stripe-ordered and superconducting copper oxides. These results imply the existence of dynamic charge inhomogeneity due to strong electronic correlations as the observed effects are absent in calculations using conventional techniques. The disappearance of the phonon anomaly at the nonsuperconducting extremes of doping suggests that electron-phonon coupling may be important to the mechanism of superconductivity, although the contribution may be indirect.

[1] Lanzara, A. et al., Evidence for ubiquitous strong electron-phonon coupling in high-temperature superconductors, Nature **412**, 510-514 (2001).

[2] Devereaux, T.P., Cuk, T., Shen, Z.-X. & Nagaosa, N. Anisotropic electron-phonon interaction in the cuprates. Phys. Rev. Lett. **93**, 117004 (2004).

[3] Bohnen, K.-P., Heid, R. & Krauss, M. Phonon dispersion and electron-phonon interaction for $YBa_2Cu_3O_7$ from first-principles calculations. Europhys. Lett. **64**, 104-110 (2003).

[4] Falter, C. & Hoffmann, G. A. Nonlocal electron-phonon coupling of ionic charge-fluctuation type and phonon anomalies in high-temperature superconductors. Phys. Rev. B **64**, 054516 (2001).

[5] Baron, A. Q. R. et al., Kohn Anomaly in $MgB_2$ by Inelastic X-Ray Scattering, Phys. Rev. Lett. **92**, 197004 (2004).

[6] Tranquada, J.M. *et al.* Evidence for stripe correlations of spins and holes in copper oxide superconductors. Nature **375**, 561-563 (1995).

[7] Machida, K., Magnetism in $La_2CuO_4$ based compounds, Physica C **158** 192-196 (1989).

[8] Zaanen, J. & Gunnarsson, Charged magnetic domain lines and the magnetism of high-$T_c$ oxides Phys. Rev. B **40** 7391-7394 (1989).

[9] Kivelson, S.A. *et al.* How to detect fluctuating stripes in the high-temperature superconductors. Rev. Mod. Phys. **75**, 1201-1241 (2003).

[10] McQueeney, *et al.* Anomalous dispersion of LO phonons in $La_{1.85}Sr_{0.15}CuO_4$ at low temperatures. Phys. Rev. Lett. **82**, 628-631 (1999).

[11] Chung, J.-H. et al. In-plane anisotropy and temperature dependence of oxygen phonon modes in YBa$_2$Cu$_3$O$_{6.95}$ Phys. Rev. B **67**, 014517 (2003).

[12] Pintschovius, L. *et al*., Lattice dynamical studies of HTSC materials, Physica C **185-189**, 156-161 (1991).

[13] Tranquada, J. M., Nakajima, K., Braden, M., Pintschovius, L., and McQueeney, R. J. Bond-stretching-phonon anomalies in stripe-ordered La$_{1.69}$Sr$_{0.31}$NiO$_4$. Phys. Rev. Lett. **88**, 075505 (2002).

[14] Ito, M. *et al.* Effects of "stripes" on the magnetic excitation spectra of La$_{1.48}$Nd$_{0.4}$Sr$_{0.12}$CuO$_4$. J. Phys. Soc. Jpn. **72**, 1627-1630 (2003).

[15] Fujita, M., Goka, H., Yamada, K., Tranquada, J. M. & Regnault, L. P. Stripe order, depinning, and fluctuations in La$_{1.875}$Ba$_{0.125}$CuO$_4$ and La$_{1.875}$Ba$_{0.075}$Sr$_{0.050}$CuO$_4$. Phys. Rev. B **70**, 104517 (2004).

[16] P. Abbamonte et al. Spatially modulated 'Mottness' in La$_{2-x}$Ba$_x$CuO$_4$, Nature Physics **1**, 155-158 (2005).

[17] Chaplot, S. L., Reichardt, W., Pintschovius, L. & Pyka, N. Common interatomic potential model for the lattice dynamics of several cuprates. Phys. Rev. B **52**, 7230-7242 (1995).

[18] Pintschovius, L., Reichardt, W., Kläser, M., Wolf, T. & v. Löhneysen, H. Pronounced in-Plane anisotropy of phonon anomalies in YBa$_2$Cu$_3$O$_{6.6}$. Phys. Rev. Lett. **89**, 037001 (2002).

[19] Pintschovius, L. *et al* Oxygen phonon branches in YBa$_2$Cu$_3$O$_7$, Phys. Rev. B **69**, 214506 (2004).

[20] Uchiyama, H. et al., "Softening of Cu-O Bond Stretching Phonons in Tetragonal HgBa$_2$Cu O$_{4+\delta}$", Phys. Rev. Lett. **92**, 197005 (2004).

[21] Pintschovius, L. *et al.*, Evidence for dynamic charge stripes in the phonons of optimally doped YBCO. http://www.arxiv.org/PS_cache/cond-mat/pdf/0308/0308357.pdf (2003)

[22] Pintschovius, L. & Braden, M. Anomalous dispersion of LO phonons in La$_{1.85}$Sr$_{0.15}$CuO$_4$. Phys. Rev. B **60**, R15039 (1999).

[23] Fukuda et al. Doping dependence of softening in the bond-stretching phonon mode of La$_{2-x}$Sr$_x$CuO$_4$ (0 ≤ x ≤ 0.29) Phys. Rev. B **71**, 060501(R) (2005).

[24] Reznik, D., Keimer, B., Dogan, F. & Aksay, I.A. *q* Dependence of self-energy effects of the plane oxygen vibration in YBa$_2$Cu$_3$O$_7$. Phys. Rev. Lett. **75**, 2396-2399 (1995).

[25] Rösch, O. & Gunnarsson, O. Electron-phonon Interaction in the *t–J* Model. Phys. Rev. Lett. **92**, 146403 (2004).

[26] Ishihara, S. & Nagaosa, N. Interplay of electron-phonon interaction and electron correlation in high-temperature superconductivity. Phys. Rev. B **69**, 144520 (2004).

[27] Kaneshita, E., Ichioka, M. & Machida, K. Phonon anomalies due to collective stripe modes in high T$_c$ cuprates. Phys. Rev. Lett. **88**, 115501 (2002).

[28] H. Hiraka et al., Spin Fluctuations in the Underdoped High-*T*$_c$ Cuprate La$_{1.93}$Sr$_{0.07}$CuO$_4$, Phys. Soc. Jpn. 70, 853-858 (2001).

[29] Hinkov, V. *et al.*, Two-dimensional geometry of spin excitations in the high temperature, superconductor YBa$_2$Cu$_3$O$_{6+x}$, Nature **430**, 650-654 (2004).

30 Zhou, X. J. *et al.*, One-Dimensional Electronic Structure and Suppression of d-Wave Node State in $La_{1.28}Nd_{0.6}Sr_{0.12}CuO_4$  Science **286**, 268-272 (1999).

**Acknowledgements** GDG and JMT are supported by the Office of Science, U.S. Department of Energy. KY, MF and MS are supported by grants from the MEXT of Japan. DR would like to thank S. Kivelson for useful comments on the first version of the manuscript.
**Author Information** The authors declare no competing financial interests.

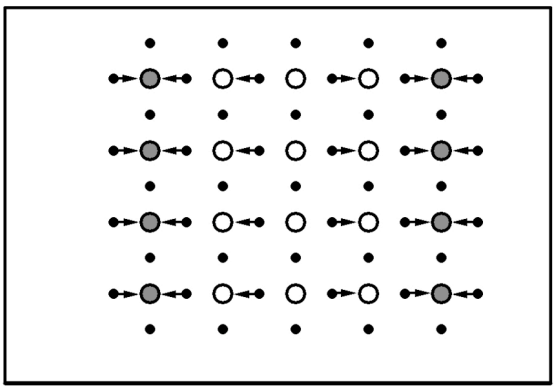

**Figure 1** Displacement pattern of the oxygen ions for the phonon with **q**=(0.25 0 0) propagating perpendicular to the stripes. Arrows represent atomic displacements. Large circles correspond to copper and small ones to oxygen. Open large circles represent hole-poor antiferromagnetic regions, while the filled circles represent the hole-rich lines.

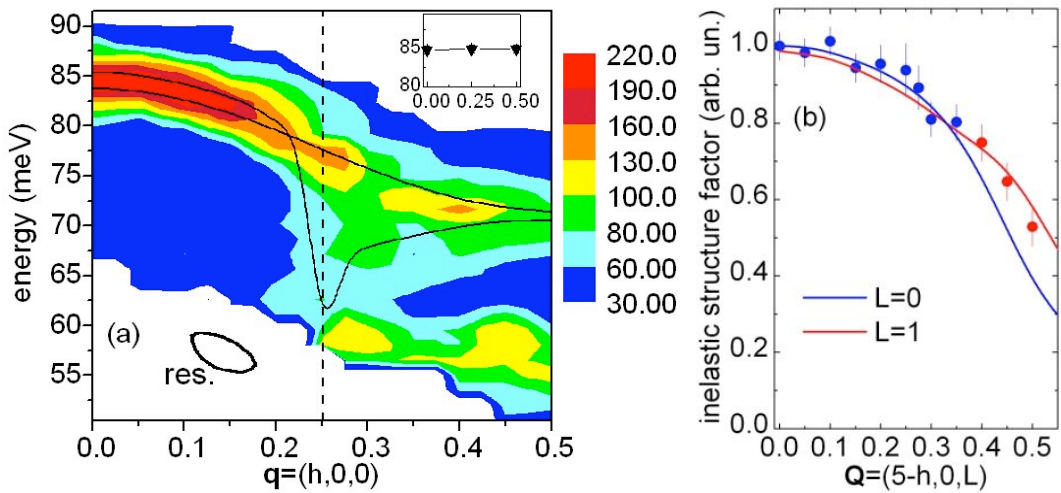

**Figure 2** Bond-stretching phonon branch in $La_{1.875}Ba_{0.125}CuO_4$. The experiments were performed on the 1T triple-axis spectrometer at the ORPHEE reactor at LLB, Saclay. A Cu220 reflection was utilized as the monochromator in order to achieve high resolution, while the 002 Pyrolytic Graphite (PG) analyzer reflection was fixed at 14.7meV. The wavevectors are given in reciprocal lattice units of ($2\pi/a$, $2\pi/b$, $2\pi/c$), where a=b=3.78 Å, and c=13.18 Å. The scattering cross section was measured in the Brillouin zone (BZ) near the reciprocal lattice vector **G**=(5 0 L) (L=0,1, or 2) to maximize accessible dynamic structure factor of the bond-stretching phonons. For neutron momentum transfer $\hbar$**Q**, **q** = **G** – **Q**. The single crystal sample of $La_{1.875}Ba_{0.125}CuO_4$[15] weighing 5g was mounted on the spectrometer with the b-axis vertical. For temperature control, each sample was mounted in a closed-cycle refrigerator. (a) Color-coded contour plot of the intensities observed on $La_{1.875}Ba_{0.125}CuO_4$ at T = 10 K. The intensities above and below 60 meV are associated with plane-polarized Cu-O bond-stretching vibrations and bond-bending vibrations, respectively. Black lines are dispersion curves evaluated from two-peak fits to the data (one of them has a cosine-behavior). The white area at the lower left corner of the diagram was not accessible in this experiment. The ellipse illustrates the instrumental resolution. The inset shows the dispersion in the [110]-direction.

The dashed line represents the charge–ordering wave vector. (b) The dots denote inelastic structure factors of the bond-stretching modes as deduced from the integrated intensities observed for wave vectors **Q** = (5-h,0,0) (blue) and **Q** = (5-h,0,L) (red); error bars represent s.d. The data were normalized to the value at **Q** = (5,0,0). The solid lines were calculated from a shell model.[17] We note that the inelastic structure factors depend very little on the value of L except for wave vectors close to the zone boundary.

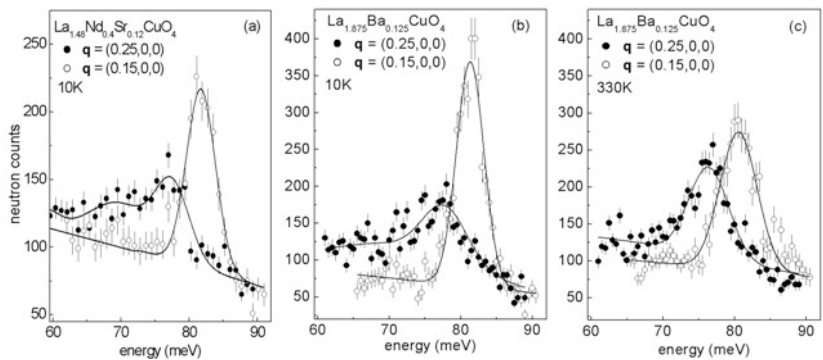

**Figure 3** Representative energy scans at 10K and 330K. Energy scans taken on $La_{1.48}Nd_{0.4}Sr_{0.12}CuO_4$ at T = 10 K (a) and on $La_{1.875}Ba_{0.125}CuO_4$ (b,c) at 10K (b) and 330K (c). The sample of $La_{1.48}Nd_{0.4}Sr_{0.12}CuO_4$[14] consisted of six single crystals coaligned with the c-axis vertical. The phonon at **q**=(0.15 0 0) is "normal", i.e. it has a gaussian lineshape on top of a linear background, which results from multiphonon and incoherent scattering and has no strong dependence on **q**. Its intensity reduction from 10K (b) to 330K (c) is consistent with the Debye-Waller factor. At **q**=(0.25 0 0), there is extra intensity on top of the background in the tail of the main peak. It results from one-phonon scattering that extends to the lowest investigated energies, while the peak intensity is greatly suppressed as discussed in the text. The effect is reduced

but does not disappear on heating from 10K to 330K. Note that **G**=(5 0 2) for (a) and (5 0 0) for (b) and (c).

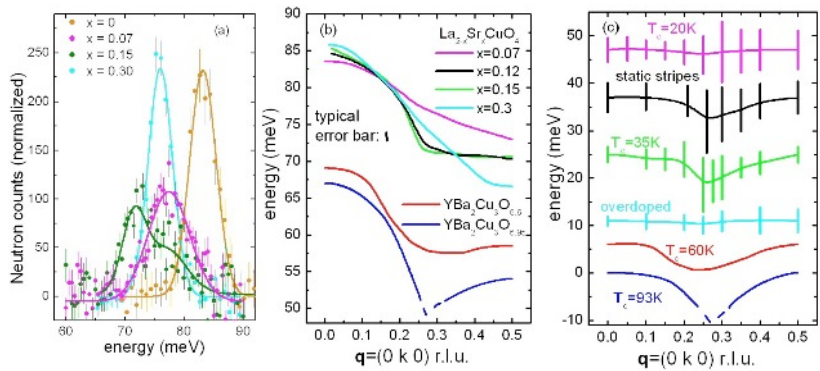

**Figure 4** Correlation of the phonon anomaly with stripe order and superconductivity. (a) Comparison of energy scans taken at 10 K on $La_{2-x}Sr_xCuO_4$ at **q** = (0.25,0,0), after subtraction of linear background. Solid lines are guides to the eye; error bars represent s.d. The resolution was somewhat lower in the measurement on the x=0 sample (L. Pintschovius, W. Reichardt, and A. Yu Rumiantsev, unpublished results), where Cu111, rather than Cu220, monochromator was used. (b) Dispersion of the anomalous branch in $La_{2-x}Sr_xCuO_4$ (x=0.07, 0.12, and 0.15) and the (0 1 0) bond-stretching branch along the chain direction in $YBa_2Cu_3O_{6.6}$ ($T_c$=66K)[18] and $YBa_2Cu_3O_{6.95}$ ($T_c$=93K)[19]. The x=0.12 compound was $La_{1.48}Nd_{0.4}Sr_{0.12}CuO_4$. The curve for $YBa_2Cu_3O_{6.95}$ is only approximate, since the anomaly is so strong near **q**=(0 0.25 0) that there is no well defined peak there (see Fig. 1 in Ref. 19). The difference in energies at q=0 is results from slightly different Cu-O bond lengths. Dashed blue line represents the region where the bond-stretching mode mixes with others and its precise energy cannot be established.[21] (c) Difference between the downward cosine dispersion, fit to the zone center and the zone boundary for each sample, and the dispersions in (b); vertical lines represent peak widths. For the

YBCO data the width information is not available because the data were taken in the defocusing scattering geometry with relatively poor energy resolution. Curves are offset for clarity. Dashed blue line is the same as in (b).